\documentclass[11pt,a4paper]{article}
\usepackage{latexsym}
\begin{document}
\setlength{\unitlength}{1mm}
\newcommand{\Section}{\setcounter{equation}{0}\section}
\renewcommand{\figurename}{Fig.}

\begin{center}
{\bf   High-Energy Antinucleus-Nucleus Collisions 

                                         and 

                 Extended Multi-Chain Model }

$\vspace{2cm}$

\footnotesize{$\rm{HUJIO\  NOD{A}^{\sharp,\P},\ SHIN\rm{-}ICHI\ NAKARIK{I}^{+,\|}\ and\ TSUTOMU\ TASHIR{O}^{+,\dagger}}$

$\sharp\ Faculty\ of\ Science,\ Ibaraki\ University,\ Mito\ 310-8512,\ Japan$\footnote{Now, Emeritus Professor of Ibaraki University}

$+\ Research\ Institute\ of\ Natural\ Science,\ Okayama\ University\ of\ Science,\ Okayama\ $

$700-0005,\ Japan$

$\P:\ noda@mcs.ibaraki.ac,jp$

$\|:\ nakariki@sp.ous.ac.jp$

$\dagger:\ tashiro@sp.ous.ac.jp$
}

\end{center}

$\vspace{0.5cm}$

\small{High-energy antinucleus-nucleus collisions are studied in the extended multi-chain model.  The event probability of inclusive process is calculated by means of the operator matrix in the moment space. Analytic forms for single-particle distribution of inclusive process are derived.}

$\vspace{0.5cm}$

\small{$Keywords$:$\ $Antinucleus-nucleus collisions;$\ $Extended multi-chain model;$\ $ Single-particle distribution }

$\vspace{0.5cm}$

\small{\section{Introduction}}

Recently,  the rapidity distributions of the charged particles and the ratios of the numbers of anti-hadrons to hadrons in nucleus-nucleus collisions have been investigated at SPS[1,2] and RHIC[3]. Also, the possibility of the formation of a quark-gluon plasma state has been studied very intensively.

The nucleon-nucleus($N$-$A$)  and nucleus-nucleus($A$-$A$) collisions have been studied from the multiple-scattering view[4,5,6]. The  basic view of the multi-chain model(MCM) was discussed in [7]. Kinoshita etal. investigated this view in detail and succeeded  phenomenologically to reproduce the old data on the multiplicity and the inclusive spectra of the leading nucleon and secondary particles  etc. in collisions of $N$-$A$[8] and $A$-$A$[9,10]. In MCM[11], it is assumed that the nucleon-nucleon($N$-$N$) interaction is considered as a exchange of one multi peripheral chain from which secondary hadrons are emitted as shown in Fig.1. However, it was difficult to perform analytical calculation for $A$-$A$ collisions because of the complexity of the general formula in resolving the full combination of chain configuration. Also, the probabilistic approach[12] and the leading cluster cascade model with a recurrence equation[13] were studied for $\bar{N}$-$A$ collisions. The essential point of MCM with successive collisions is that the number of $N$-$N$ collision is equal to the mass number of nucleus A. Namely, nucleus looks as if nucleons in nucleus stand in a line.

\setlength{\unitlength}{1mm}
\begin{figure}
\begin{center}
\begin{picture}(120,40)
\thicklines

\put(30,10){\line(60,0){60}}
\put(30,15){\line(60,0){60}}
\put(30,20){\line(60,0){60}}
\put(30,40){\line(60,0){60}}
\put(30,45){\line(60,0){60}}

\put(40,45){\line(-2,1){4}}
\put(40,45){\line(-2,-1){4}}
\put(80,45){\line(-2,1){4}}
\put(80,45){\line(-2,-1){4}}

\put(40,40){\line(-2,1){4}}
\put(40,40){\line(-2,-1){4}}
\put(80,40){\line(-2,1){4}}
\put(80,40){\line(-2,-1){4}}

\put(40,20){\line(-2,1){4}}
\put(40,20){\line(-2,-1){4}}
\put(40,15){\line(-2,1){4}}
\put(40,15){\line(-2,-1){4}}
\put(40,10){\line(-2,1){4}}
\put(40,10){\line(-2,-1){4}}

\put(80,20){\line(-2,1){4}}
\put(80,20){\line(-2,-1){4}}
\put(80,15){\line(-2,1){4}}
\put(80,15){\line(-2,-1){4}}
\put(80,10){\line(-2,1){4}}
\put(80,10){\line(-2,-1){4}}

\put(50,20){\circle*{1.4}}
\put(50,40){\circle*{1.4}}
\put(55,10){\circle*{1.4}}
\put(55,45){\circle*{1.4}}
\put(65,40){\circle*{1.4}}
\put(65,15){\circle*{1.4}}

\put(50,21){\oval(2,2)[r]}
\put(50,23){\oval(2,2)[l]}
\put(50,25){\oval(2,2)[r]}
\put(50,27){\oval(2,2)[l]}
\put(50,29){\oval(2,2)[r]}
\put(50,31){\oval(2,2)[l]}
\put(50,33){\oval(2,2)[r]}
\put(50,35){\oval(2,2)[l]}
\put(50,37){\oval(2,2)[r]}
\put(50,39){\oval(2,2)[l]}

\put(65,16){\oval(2,2)[r]}
\put(65,18){\oval(2,2)[l]}
\put(65,20){\oval(2,2)[r]}
\put(65,22){\oval(2,2)[l]}
\put(65,24){\oval(2,2)[r]}
\put(65,26){\oval(2,2)[l]}
\put(65,28){\oval(2,2)[r]}
\put(65,30){\oval(2,2)[l]}
\put(65,32){\oval(2,2)[r]}
\put(65,34){\oval(2,2)[l]}
\put(65,36){\oval(2,2)[r]}
\put(65,38){\oval(2,2.5)[l]}

\put(55,12){\oval(2,2)[r]}
\put(55,14){\oval(2,2)[l]}
\put(55,16){\oval(2,2)[r]}
\put(55,18){\oval(2,2)[l]}
\put(55,20){\oval(2,2)[r]}
\put(55,22){\oval(2,2)[l]}
\put(55,24){\oval(2,2)[r]}
\put(55,26){\oval(2,2)[l]}
\put(55,28){\oval(2,2)[r]}
\put(55,30){\oval(2,2)[l]}
\put(55,32){\oval(2,2)[r]}
\put(55,34){\oval(2,2)[l]}
\put(55,36){\oval(2,2)[r]}
\put(55,38){\oval(2,2)[l]}
\put(55,40){\oval(2,2)[r]}
\put(55,43){\oval(2,3)[l]}

\put(25,42.5){A}
\put(25,15){B}

\end{picture}
\end{center}
\caption{ MCM for $A$-$B$ collision with $A=2$ and $B=3$. Wavy lines represents the inelastic $N$-$N$ interaction, namely, $N+N \to N+N$+hadrons.}

\end{figure}
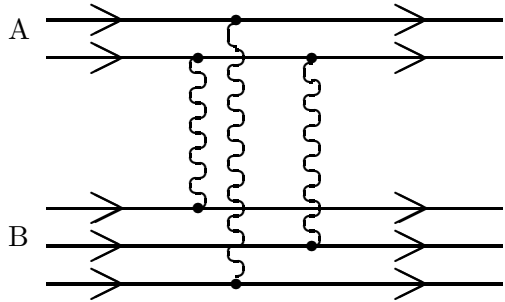

In previous paper[14], we investigated the extended MCM(EMCM) on the basis of MCM with successive collision[15] and estimated the full combinations of chain configuration in terms of the vector-operator method in the moment space by consulting the Mellin transformation[13]. We derived the analytic forms for single-particle distributions in $A$-$A$ collisions.  This approach may apply to the $\bar{N}$-$A$, but does not extend to $\bar{A}$-$A$ collisions.  

In this paper, we propose the new EMCM, namely, a unified treatment of  $N$-$A$, $\bar{N}$-$A$, $A$-$A$ and $\bar{A}$-$A$ collisions. We introduce the concept of the operator matrix in the moment space to characterize these collisions. By means of the operator matrix, we calculate the single-particle distribution of the inclusive process $A+B \to C+X $. 
We assume the dynamics to scale with energy.  A nucleus is treated as a set of mutually independent nucleons. The cascading of the produced hadrons is neglected owing to the long formation length in nucleus. Also, the longitudinal motion is treated. The transverse momentum distribution is assumed to be independent of  incident energy and longitudinal momentum.

In Section 2, the EMCM are investigated and the vector-operator formalism in the moment space is given. The operator matrix is introduced to treat the $A$-$A$ collision and the calculation method is given. 
In Section 3, the structure of $\bar{N}$-$N$ collision is studied.  The single-particle distributions of $\bar{N}$-$A$ collisions are investigated and their analytic forms are derived. In Section 4, $\bar{A}$-$A$ collisions are studied. 
 In Section 5, conclusion and discussion are given. The definition and properties of operator matrix are given in Appendix.

\vspace{0.5cm}

\small{\section{Extended multi-chain model}} 

\small{\subsection{Moment space and vector-operator notation}}

We consider $A$-$B$ collisions. The mass numbers of the projectile nucleus A and target nucleus B are $A$ and $B$, respectively.  We pay attention to the single-particle distribution of the inclusive process $A+B \to C+X(\rho^{AB}_C(x)) $ defined by 
\begin{eqnarray}
\rho ^{AB}_C(x)=x\frac{d\sigma (A+B \to C+X)}{dx},
\end{eqnarray}
where $C$ is the produced hadron and $x$ denotes the longitudinal momentum fraction(Feynman variable). We express the projectile(target) fragmentation regions as $x>0\ (\ x<0\ )$.

We assume that a nucleus is treated as a set of mutually independent nucleons.
By the Mellin transformation of Eq.(1), the moment $\rho^{AB}_C(\alpha)$ is defined by
\begin{eqnarray}
\rho^{AB}_C (\alpha)=\int_0^1 dxx^{\alpha-2}\rho^{AB}_C (x).
\end{eqnarray}
The inverse Mellin transformation is defined by $\rho ^{AB}_C(x)=\frac{1}{2\pi i}\int_{\tau-i\infty}^{\tau+i\infty}d\alpha x^{1-\alpha} \rho^{AB}_C(\alpha)$.
 Here, we employ a vector-operator notation[15]. The operator $\hat {\rho}^{AB}(\alpha)$ in the moment space is defined by 
\[ <C|\hat{\rho}^{AB}(\alpha)|N>=\rho^{AB}_C(\alpha),\]
where $<C|$ denotes the produced hadrons in the final state and $|N>$ denotes the nucleon state in the projectile nucleus A or the target nucleus B.

\vspace{0.5cm}

\small{\subsection{$A$-$B$ collisions and operator matrix $\hat{\mbox {\boldmath $Q$}}_{AB}(\vec{b},\alpha)$}} 

The $A$-$B$ collision is characterized by the $N$-$N$ collisions between the nucleon in nucleus A and the nucleon in nucleus B. The peripheral reaction $N+N \to N+N$+hadrons is considered where $N$ in the final state denotes the leading nucleon. In the MCM with successive collisions, the nucleon in nucleus A successively collides with the nucleons in nucleus B.Then, the collison number of times is limited to $B$ and vice versa. Thus, the total collision number in $A$-$B$ collision is $AB$. 

We treat the $A$-$B$ collision at the impact parameter $\vec{b}$ and in the moment space $\alpha$. We introduce the operator $\hat{Q}_{ij}(\vec{b}, \alpha) $ to denote the probability in the impact parameter $\vec{b}$ and the longitudinal momentum fraction distribution in the moment space of the collision between the $i$-th nucleon in A and the $j$-th nucleon in B where $i=1,2,\ldots,A$ and $j=1,2,\ldots,B$.
The operator matrix $\hat{\mbox {\boldmath $Q$}}_{AB}(\vec{b},\alpha)$ to characterize the total structure of $A$-$B$ collisions, is defined by the rectangular or square array of the operators $\hat{Q}_{ij}(\vec{b}, \alpha) $ with $A$ rows and $B$ columns 
\begin{eqnarray}
\hat{\mbox {\boldmath $Q$}}_{AB}(\vec{b}, \alpha)=\left[ \matrix{
                                                \hat{Q}_{11}(\vec{b}, \alpha) &  \hat{Q}_{12}(\vec{b}, \alpha) &  \cdots &  \hat{Q}_{1B}(\vec{b}, \alpha) \cr
                                                        \vdots & \vdots & \ddots & \vdots \cr
                                                 \hat{Q}_{A1}(\vec{b}, \alpha) &  \hat{Q}_{A2}(\vec{b}, \alpha) &  \cdots &  \hat{Q}_{AB}(\vec{b}, \alpha) 
                                                 } \right].
\end{eqnarray}
By the means of the calculus of operator matrix  given in Appendix, we may derive the informations on the single-particle distributions of $A+B \to C+X$ by the sum decomposition rules defined in Appendix from $\hat{\mbox {\boldmath $Q$}}_{AB}(\vec{b},\alpha)$. Namely, we use the sum of the products of the row elements(${\bf Spr}$) and the sum of the products of the column elements(${\bf Spc}$) for $\hat{\mbox {\boldmath $Q$}}_{AB}(\vec{b},\alpha)$ in regions of $x>0$ and $x<0$, respectively. The calculus of operator matrix  gives the convenience of calculating the probability of one event in $A$-$B$ collision.
For example, this operator matrix of the $A$-$B$ collision with $A=3$ and $B=4$ is shown by the diagram in Fig.2 in stead of Fig.1. The intersection point in the diagram denotes the $N$-$N$ collision between the nucleon in A and the nucleon in B. It is assumed that the time of the $i$-th nucleon in A and the $j$-th evolve from $\hat{Q}_{iB}(\vec{b}, \alpha) $ to $\hat{Q}_{i1}(\vec{b}, \alpha) $ and from $\hat{Q}_{Aj}(\vec{b}, \alpha) $ to $\hat{Q}_{1j}(\vec{b}, \alpha) $ in Fig.2, respectively.  The ordering of the operator $\hat{Q}_{ij}(\vec{b}, \alpha)$ obeys this time-evolution.

\setlength{\unitlength}{1mm}
\begin{figure}
\begin{center}
\begin{picture}(120,40)
\thicklines

\put(30,35){\line(50,0){50}}
\put(30,30){\line(50,0){50}}
\put(30,25){\line(50,0){50}}

\put(45,10){\line(0,40){40}}
\put(50,10){\line(0,40){40}}
\put(55,10){\line(0,40){40}}
\put(60,10){\line(0,40){40}}

\put(35,35){\line(2,1){4}}
\put(35,35){\line(2,-1){4}}
\put(70,35){\line(2,1){4}}
\put(70,35){\line(2,-1){4}}

\put(35,30){\line(2,1){4}}
\put(35,30){\line(2,-1){4}}
\put(70,30){\line(2,1){4}}
\put(70,30){\line(2,-1){4}}

\put(35,25){\line(2,1){4}}
\put(35,25){\line(2,-1){4}}
\put(70,25){\line(2,1){4}}
\put(70,25){\line(2,-1){4}}

\put(45,17.5){\line(1,-2){2}}
\put(45,17.5){\line(-1,-2){2}}
\put(45,45){\line(1,-2){2}}
\put(45,45){\line(-1,-2){2}}

\put(50,17.5){\line(1,-2){2}}
\put(50,17.5){\line(-1,-2){2}}
\put(50,45){\line(1,-2){2}}
\put(50,45){\line(-1,-2){2}}

\put(55,17.5){\line(1,-2){2}}
\put(55,17.5){\line(-1,-2){2}}
\put(55,45){\line(1,-2){2}}
\put(55,45){\line(-1,-2){2}}

\put(60,17.5){\line(1,-2){2}}
\put(60,17.5){\line(-1,-2){2}}
\put(60,45){\line(1,-2){2}}
\put(60,45){\line(-1,-2){2}}

\thicklines
\put(45,35){\circle*{1.4}}
\put(50,35){\circle*{1.4}}
\put(55,35){\circle*{1.4}}
\put(60,35){\circle*{1.4}}

\put(45,30){\circle*{1.4}}
\put(50,30){\circle*{1.4}}
\put(55,30){\circle*{1.4}}
\put(60,30){\circle*{1.4}}

\put(45,25){\circle*{1.4}}
\put(50,25){\circle*{1.4}}
\put(55,25){\circle*{1.4}}
\put(60,25){\circle*{1.4}}

\put(65,35.8){1}
\put(65,30.8){2}
\put(65,25.8){3}

\put(47.0,20){1}
\put(52.0,20){2}
\put(57.0,20){3}
\put(62.0,20){4}

\put(25,35){N}
\put(25,30){N}
\put(25,35){N}

\put(42.5,51){N}
\put(47.5,51){N}
\put(52.5,51){N}
\put(57.5,51){N}

\put(80,30){\oval(2,18)}
\put(52.5,10){\oval(25,2.5)}

\put(81,31){\line(3,0){3}}
\put(81,29){\line(3,0){3}}
\put(51.5,8.75){\line(0,-3){3}}
\put(53.5,8.75){\line(0,-3){3}}

\put(63,51){$(x<0)$}
\put(22.5,20){$(x>0)$}

\put(85,30){A}
\put(52.5,2.5){B}

\end{picture}
\end{center}
\caption{ EMCM for $A$-$B$ collision with $A=3$ and $B=4$. The intersection points represent $N$-$N$ collision($\bullet$). The solid lines denote nucleon($N$).}
\end{figure}
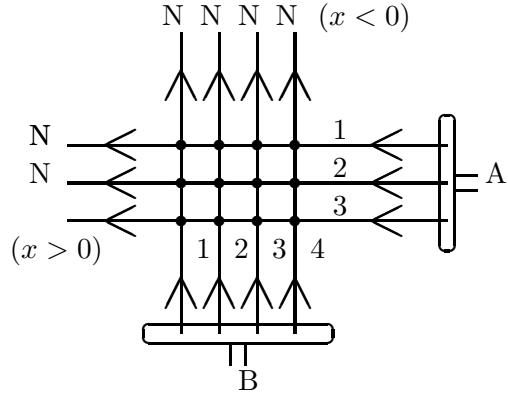

\vspace{0.5cm}

 \small{\subsection{Single-particle distribution in $A$-$B$ collisions}} 

We consider the inclusive reaction $A+B \to N+X$ where $N$ denotes the leading nucleon. 
We introduce the operator $\hat{Q}_{0}(\vec{b},\alpha)$ defined by
\begin{eqnarray}
\hat{Q}_{0}(\vec{b},\alpha)=\eta(\vec{b})\hat{G(\alpha)}+\lambda(\vec{b})\hat{J}(\alpha).
\end{eqnarray}
for $N$-$N$ collisions in nucleus. Here,  $\lambda(\vec{b})=\sigma^{NN}_{inel}\int d\vec{s} \int d\vec{t} T_A(\vec{s})T_B(\vec{t})\delta^2(\vec{b}+\vec{s}-\vec{t})$ where $\sigma^{NN}_{inel}$ is the inelastic cross section of $N$-$N$ collisions. Also, $T_A(\vec{s})$ and $T_{B}(\vec{t})$ are the normalized nuclear thickness functions of nucleus A and nucleus B, respectively. The $\lambda(\vec{b})$ denotes the inelastic interaction probability while the $\eta(\vec{b})$  expresses the passing-through probability($\eta(\vec{b})+\lambda(\vec{b})=1$ owing to the probability conservation). Operators $\hat{G}(\alpha)$ and $\hat{J}(\alpha)$ characterize the momentum fraction distributions as defined $[\hat{G}(\alpha), \hat{J}(\alpha)]=0$ and $\hat{G}(\alpha)=1$. The operator $\hat{J}(\alpha)$ expresses the inelastic interaction of $N$-$N$collisions.
By considering both operators $\hat{G}(\alpha)$ and $\hat{J}(\alpha)$, we may calculate full combinations of the chain configurations with ease. In fact,  it corresponds to MCM with two kinds of chain.

In this case, the elements of the operator matrix with $A$ rows and $B$ columns are given by $\hat{Q}_{ij}(\vec{b},\alpha)=\hat{Q}_0(\vec{b},\alpha)$ where $i=1,2,\ldots,A$ and $j=1,2,\ldots,B$, and hence
\begin{eqnarray}
\hat{\mbox {\boldmath $Q$}}_{AB}(\vec{b}, \alpha)=\left[
                                                \matrix{
                                                \hat{Q}_{0}(\vec{b}, \alpha) &  \hat{Q}_{0}(\vec{b}, \alpha) &  \cdots &  \hat{Q}_{0}(\vec{b}, \alpha) \cr
                                                        \vdots & \vdots & \ddots & \vdots \cr
                                                 \hat{Q}_{0}(\vec{b}, \alpha) &  \hat{Q}_{0}(\vec{b}, \alpha) &  \cdots &  \hat{Q}_{0}(\vec{b}, \alpha) \cr     
                                                 }
                                                 \right] .   
\end{eqnarray}

We pay attention to the projectile fragmentation regions($x>0$) and thus sum the products of the row elements of Eq.(5).
>From Eq.(56) in Appendix, we get 
\begin{eqnarray}
{\bf Spr} \hat{\mbox {\boldmath $Q$}}_{AB}(\vec{b},\alpha)=A(\eta(\vec{b})+\lambda(\vec{b}))^{AB-B}(\hat{Q}_{0}(\vec{b},\alpha))^B.
\end{eqnarray}
Furthermore, we use the relation $[\hat{G}(\alpha),\hat{J}(\alpha)]=0$. Eq.(6) reduces to
\begin{eqnarray*}
{\bf Spr} \hat{\mbox {\boldmath $Q$}}_{AB}(\vec{b},\alpha)=A\sum_{k=0}^{AB-B}\left(\matrix{ AB-B \cr
 k \cr}\right) \eta(\vec{b})^{AB-B-k}\lambda(\vec{b})^k
\end{eqnarray*}
\begin{eqnarray*}
\times \sum_{l=0}^{B}\left(\matrix{ B \cr
 l \cr}\right) \eta(\vec{b})^{B-l}\lambda(\vec{b})^l(\hat{G}(\alpha))^{B-l}(\hat{J}(\alpha))^l,
\end{eqnarray*}
\begin{eqnarray}
=A\sum_{m=0}^{AB}P_{m}(AB;\vec{b})\sum_{l=0}^{B}H(l;AB,m,B)(\hat{G}(\alpha))^{B-l}(\hat{J}(\alpha))^l,
\end{eqnarray}
where $P_m(AB;\vec{b})$ is the Glauber probability and $H(l;AB,m,B)$ is the hypergeometric distribution. They are  defined by
\[P_m(AB;\vec{b})=\left(\matrix{ AB \cr
 m \cr}\right) \eta(\vec{b})^{AB-m}\lambda(\vec{b})^m,\ \ 
H(l;AB,m,B)=\frac{\left(\matrix{ AB-m \cr
 B-l \cr}\right) 
 \left(\matrix{m \cr
 l \cr }\right)}
 {\left(\matrix{ AB \cr
 B \cr}\right)}.
\]
>From the Glauber probability, we obtain the inelastic cross section of $A$-$B$ collision given by \[\sigma_{inel}^{AB}=\int d\vec{b}\sum_{m=1}^{AB}P_m(AB;\vec{b})=\int d\vec{b}[1-(1-\lambda(\vec{b}))^{AB}]\]
 and the averaged collision number $\bar{n}=AB\sigma_{inel}^{NN}/\sigma_{inel}^{AB}$.

The matrix elements of the leading nucleon $N$ in the final state, is defined by
\begin{eqnarray}
<N|\hat{J}(\alpha)|N>=\frac{1}{\sigma^{NN}_{inel}}<N|\hat{\rho}^{NN}(\alpha)|N> \equiv F(\alpha),
\end{eqnarray}
where $<N|$ and $|N>$ denote the leading nucleon state and the nucleon state in the initial nucleus A or B.
By the inverse Mellin transformation, the fraction functions $F(x)$ is  given by
\begin{eqnarray}
F(x)=\frac{1}{2\pi i}\int_{\tau-i\infty}^{\tau+i\infty}d\alpha x^{1-\alpha}F(\alpha),
\end{eqnarray}
where $F(x)$ is normalized to unity, namely, $\int _0^1 \frac{dx}{x}F(x)=1$.

The distribution $Q^{AB}_{N/N}(\vec{b},\alpha,A)$  in the projectile fragmentation regions(A side) is defined by 
\begin{eqnarray}
Q^{AB}_{N/N}(\vec{b},\alpha,A)=<N|{\bf Spr} \hat{\mbox {\boldmath $Q$}}_{AB}(\vec{b},\alpha)|N>.
\end{eqnarray}
When we put $\hat{G}(\alpha)=1$, from Eqs.(7), (8) and (10), we get
\begin{eqnarray}
      Q^{AB}_{N/N}(\vec{b},\alpha,A) = A\sum_{m=0}^{AB}P_{m}(AB;\vec{b})\sum_{l=0}^{B}H(l;AB,m,B)F(\alpha)^l.
\end{eqnarray}

By means of the inverse Mellin transformation of Eq.(11) except for the passing-through terms, the single-particle distribution of $A+B \to N+X$  for the projectile fragmentation regions($x>0$) is given by
\begin{eqnarray}
\rho^{AB}_{N}(x)=\int d\vec{b}A\sum_{m=1}^{AB}P_{m}(AB;\vec{b})\sum_{l=1}^{B}H(l;AB,m,B)F^{(l)}(x)
\end{eqnarray}
with
\begin{eqnarray}
F^{(l)}(x)=\int_{x}^{1} \frac{dy}{y}F(\frac{x}{y})F^{(l-1)}(y) \  {\rm with}\  F^{(0)}(y)=\delta(1-y).
\end{eqnarray}
Eq.(12) agrees with the result discussed in [14]. From Eq.(12), we get the inelastic cross section given by $\sigma_{inel}^{AB}=\frac{1}{A}\int \frac{dx}{x}\rho_N^{AB}(x)=\int d\vec{b}[1-(1-\lambda(\vec{b}))^{AB}]$. Similarly, we may investigate the $N$-$A$ collisions and get the same result as the one in [14].

\vspace{0.5cm}

\small{\section{Single-particle distributions in $\bar{N}$-$A$ collisions}}

\small{\subsection{Structure of $\bar{N}$-$N$ collision in nucleus}} 

We consider the $\bar{N}$-$N$ collision in nucleus. In this case, we must introduce the annihilation interaction between $\bar{N}$ and $N$. Then, the leading meson($M_L$) is produced by the annihilation interaction as $\bar{N}+N \to M_L+M_L$+hadrons. Thus, we newly consider $M_L$-$N(\bar{N})$ collisions and $M_L$-$M_L$ collisions in nucleus. We consider the peripheral reactions $M_L+N(\bar{N}) \to M_L+N(\bar{N})$+hadrons and $M_L+M_L \to M_L+M_L$+hadrons, but neglect the non-peripheral reaction $M_L+N(\bar{N}) \to N(\bar{N})+M_L$+hadrons because of its small probability. We introduce the following operator set:

(i) $\bar{N}$-$N$ collision

\ \ (a) $\hat{Q}_n(\vec{b},\alpha)=\eta_n(\vec{b})\hat{G}_n(\alpha)+\lambda_n(\vec{b})\hat{J}_n(\alpha)$ for non-annihilation collision,

\ \ (b) $\lambda_a(\vec{b})\hat{A}(\alpha)$ for annihilation interaction,

where $\eta(\vec{b})_n+\lambda(\vec{b})_n+\lambda_a(\vec{b})=1$. Also, $\int d\vec{b} \lambda_n({\vec{b}})=\sigma_{non}^{\bar{N}N}$ and $\int d\vec{b} \lambda_a({\vec{b}})=\sigma_{an}^{\bar{N}N}$ 

where $\sigma_{non}^{\bar{N}N}$ is the non-annihilation cross section and $\sigma_{an}^{\bar{N}N}$ is the annihilation 

cross section in $\bar{N}$-$N$ collision.

(ii) $M_L$-$N(\bar{N})$ collision

\ \ (c) $\hat{Q}_L(\vec{b},\alpha)=\eta_L(\vec{b})\hat{G}_L(\alpha)+\lambda_L(\vec{b})\hat{J}_L(\alpha)$ for inelastic collision,

where $\eta(\vec{b})_L+\lambda(\vec{b})_L=1$ and $\int d\vec{b} \lambda_L({\vec{b}})=\sigma_{inel}^{M_LN(\bar{N})} .$

(ii) $M_L$-$M_L$ collision

\ \ (d) $\hat{M}_0(\vec{b},\alpha)=\eta_m(\vec{b})\hat{G}_m(\alpha)+\lambda_m(\vec{b})\hat{J}_m(\alpha)$ for inelastic collision,

where $\eta(\vec{b})_m+\lambda(\vec{b})_m=1$ and $ \int d\vec{b} \lambda_m({\vec{b}})=\sigma_{inel}^{M_LM_L}.$

The operators $\hat{G}_n(\alpha)$, $\hat{G}_L(\alpha)$ and $\hat{G}_m(\alpha)$ denote the passing-through. The annihilation operator $\hat{A}(\alpha)$ is not commutable with $\hat{Q}_n(\vec{b},\alpha)$ and $\hat{Q}_L(\vec{b},\alpha)$. The operator ordering of $\hat{A}(\alpha)$ is in front of $\hat{Q}_L(\vec{b},\alpha)$ and at the back of $\hat{Q}_n(\vec{b},\alpha)$. Notice that $[\hat{G}_n(\alpha), \hat{J}_n(\alpha)]=0$, $[\hat{G}_L(\alpha), \hat{J}_L(\alpha)]=0$ and $[\hat{G}_m(\alpha), \hat{J}_m(\alpha)]=0$. 
Also, we define 
\[<\bar{N}|\hat{J}_n(\alpha)|\bar{N}>= <N|\hat{J}_n(\alpha)|N>=\bar{F}(\alpha),\]
\[<M_L|\hat{A}(\alpha)|\bar{N}>= <M_L|\hat{A}(\alpha)|N>=A(\alpha), \]
\[<\bar{N}|\hat{J}_L(\alpha)|\bar{N}>= <N|\hat{J}_L(\alpha)|N>=N(\alpha), \]
\[<M_L|\hat{J}_L(\alpha)|M_L>= L(\alpha), \ \ <M_L|\hat{J}_m(\alpha)|M_L>=M_m(\alpha)\]
and the others of matrix element are zero. These functions in $x$ space are normalized to unity as $F(x)$.

\vspace{0.5cm}

\small{\subsection{$\bar{N}$-$A$ collisions}}

We investigate $\bar{N}$-$A$ collisions at the impact parameter $\vec{b}$. We treat the inclusive processes of $\bar{N}+A \to \bar{N}+X$ and $M_L+X$  for the projectile fragmentation regions of $\bar{N}$($x>0$) and the inclusive processes of $\bar{N}+A \to N+X$ and $M_L+X$  for the target fragmentation regions of $A$($x<0$). The $\bar{N}$-$N$ and $M_L$-$N(\bar{N})$ collisions in the operator set contribute to $\bar{N}$-$A$ collisions. 
 The operator matrix $ \hat{\mbox {\boldmath $Q$}}_{\bar{N}A}(\vec{b},\alpha)$ with one row and $A$ columns to characterize $\bar{N}$-$A$ collisions at the impact parameter $\vec{b}$ and in the moment space, is given by
  \begin{eqnarray}
\hat{\mbox {\boldmath $Q$}}_{\bar{N}A}(\vec{b},\alpha)=\hat{\mbox {\boldmath $Q$}}_{\bar{N}A}^{{\rm I}}(\vec{b},\alpha)+\hat{\mbox {\boldmath $Q$}}_{\bar{N}A}^{{\rm II}}(\vec{b},\alpha),
\end{eqnarray}
where $\hat{\mbox {\boldmath $Q$}}_{\bar{N}A}^{{\rm I}}(\vec{b},\alpha)$ and $\hat{\mbox {\boldmath $Q$}}_{\bar{N}A}^{{\rm II}}(\vec{b},\alpha)$ denote the non-annihilation and annihilation terms, respectively. They are given by
\begin{eqnarray}
\hat{\mbox {\boldmath $Q$}}_{\bar{N}A}^{{\rm I}}(\vec{b},\alpha)=[\underbrace{\hat{Q}_n(\vec{b},\alpha), \cdots, \hat{Q}_n(\vec{b})}_{A}],
\end{eqnarray}
\begin{eqnarray}
\hat{\mbox {\boldmath $Q$}}_{\bar{N}A}^{{\rm II}}(\vec{b},\alpha)=\sum_{l=0}^{A-1}\lambda_a(\hat{b})[\underbrace{\hat{Q}_L(\vec{b},\alpha), \cdots, \hat{Q}_L(\vec{b}, \alpha)}_{l}, \hat{A}(\alpha), \underbrace{\hat{Q}_n(\vec{b},\alpha), \cdots, \hat{Q}_n(\vec{b},\alpha)}_{A-l-1}].
\end{eqnarray}
The diagram for the annihilation term in $\bar{N}$-$A$ collision is shown in Fig.3.

\setlength{\unitlength}{1mm}
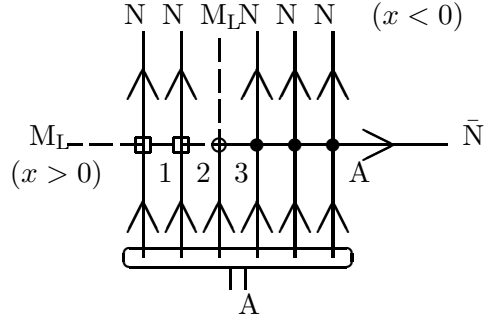
\begin{figure}
\begin{center}
\begin{picture}(120,40)
\thicklines

\put(50,25){\line(30,0){30}}

\put(45,10){\line(0,30){30}}
\put(50,10){\line(0,15){15}}
\put(55,10){\line(0,30){30}}
\put(60,10){\line(0,30){30}}

\put(73,25){\line(-2,1){4}}
\put(73,25){\line(-2,-1){4}}

\put(45,17.5){\line(1,-2){2}}
\put(45,17.5){\line(-1,-2){2}}
\put(45,35){\line(1,-2){2}}
\put(45,35){\line(-1,-2){2}}

\put(50,17.5){\line(1,-2){2}}
\put(50,17.5){\line(-1,-2){2}}

\put(65,17.5){\line(1,-2){2}}
\put(65,17.5){\line(-1,-2){2}}
\put(40,17.5){\line(1,-2){2}}
\put(40,17.5){\line(-1,-2){2}}

\put(65,35){\line(1,-2){2}}
\put(65,35){\line(-1,-2){2}}
\put(40,35){\line(1,-2){2}}
\put(40,35){\line(-1,-2){2}}

\put(55,17.5){\line(1,-2){2}}
\put(55,17.5){\line(-1,-2){2}}
\put(55,35){\line(1,-2){2}}
\put(55,35){\line(-1,-2){2}}

\put(60,17.5){\line(1,-2){2}}
\put(60,17.5){\line(-1,-2){2}}
\put(60,35){\line(1,-2){2}}
\put(60,35){\line(-1,-2){2}}

\put(30,25){\line(2,0){2}}
\put(33,25){\line(2,0){2}}
\put(34,25){\line(2,0){2}}
\put(37,25){\line(2,0){2}}
\put(38,25){\line(2,0){2}}
\put(41,25){\line(2,0){2}}
\put(42,25){\line(2,0){2}}
\put(45,25){\line(2,0){2}}
\put(46,25){\line(2,0){2}}
\put(49,25){\line(2,0){2}}

\put(50,25){\line(0,2){2}}
\put(50,28){\line(0,2){2}}
\put(50,29){\line(0,2){2}}
\put(50,32){\line(0,2){2}}
\put(50,33){\line(0,2){2}}
\put(50,36){\line(0,2){2}}
\put(50,37){\line(0,2){2}}

\put(44,24){\line(0,2){2}}
\put(46,24){\line(0,2){2}}
\put(44,24){\line(2,0){2}}
\put(44,26){\line(2,0){2}}

\put(39,24){\line(0,2){2}}
\put(41,24){\line(0,2){2}}
\put(39,24){\line(2,0){2}}
\put(39,26){\line(2,0){2}}

\put(50,25){\circle{1.8}}

\put(55,25){\circle*{1.8}}
\put(60,25){\circle*{1.8}}
\put(65,25){\circle*{1.8}}

\put(42,20){1}
\put(47.0,20){2}
\put(52.0,20){3}

\put(67.0,20){A}

\put(25,25){${\rm M_L}$}

\put(37.5,41){N}
\put(42.5,41){N}
\put(47.5,41){${\rm M_L}$}
\put(52.5,41){N}
\put(57.5,41){N}
\put(62.5,41){N}

\put(52.5,10){\oval(30,2.5)}

\put(51.5,8.75){\line(0,-3){3}}
\put(53.5,8.75){\line(0,-3){3}}

\put(70,41){$(x<0)$}
\put(22.5,20){$(x>0)$}

\put(82,25){$\bar{{\rm N}}$}
\put(52.5,2.5){A}

\put(40,10){\line(0,30){30}}
\put(65,10){\line(0,30){30}}

\end{picture}
\end{center}
\caption{ Diagram for the annihilation term in $\bar{N}$-$A$ collision. The intersection points represent $M_L$-$N$ collision($\Box$), $\bar{N}$-$N$ collision($\bullet$) and annihilation interaction of $\bar{N}$-$N$ collision($\circ$).  The solid and dashed lines denote nucleon($N$) and leading meson($M_L$), respectively.}
\end{figure}

(1) $\bar{N}+A \to \bar{N}+X$ and $M_L+X$  for the projectile fragmentation regions($x>0$)

\vspace{0.3cm}

We apply the sum decomposition rule of Eq.(52) in Appendix to $\hat{Q}_{\bar{N}A}(\vec{b},\alpha)$.
>From Eq.(15), we have
\begin{eqnarray}
{\bf Spr} \hat{\mbox {\boldmath $Q$}}_{\bar{N}A}^{{\rm I}}(\vec{b},\alpha)=(\hat{Q}_n(\vec{b},\alpha))^A=\sum_{m=0}^{A}\bar{P}_m(A;\vec{b})(\hat{J}_n(\alpha))^m,
\end{eqnarray}
where $\bar{P}_m(A;\vec{b})=\left(\matrix{ A \cr
 m \cr}\right)\eta_n(\vec{b})^{A-m}\lambda_n({\vec{b})^m}$ and then $\hat{G}(\alpha)=1.$
 
 From Eq.(16), we have
\begin{eqnarray*}
{\bf Spr} \hat{\mbox {\boldmath $Q$}}_{\bar{N}A}^{{\rm II}}(\vec{b},\alpha)=\sum_{l=0}^{A-1}\lambda_a(\vec{b})(\hat{Q}_L(\vec{b},\alpha))^l\hat{A}(\alpha)(\hat{Q}_n(\vec{b},\alpha))^{A-l-1}
\end{eqnarray*}
\begin{eqnarray}
=\sum_{l=0}^{A-1}\sum_{s=0}^{l}\sum_{t=0}^{A-l-1}\lambda_a(\vec{b})P_{s}^{L}(l;\vec{b})\bar{P}_{t}(A-l-1;\vec{b})(\hat{J}_L(\alpha))^{s}\hat{A}(\alpha)(\hat{J}_n(\alpha))^t,
\end{eqnarray}
where $P_s^L(l;\vec{b})=\left(\matrix{ l \cr
 s \cr}\right)\eta_L(\vec{b})^{l-s}\lambda_L({\vec{b})^s}$ and then $\hat{G}_L(\alpha)=\hat{G}_n(\alpha)=1.$

Taking the $<\bar{N}|\ldots |\bar{N}>$ matrix element of Eq.(17), we obtain
\begin{eqnarray}
\rho_{\bar{N}}^{\bar{N}A}(x)=\int d\vec{b}\sum_{m=1}^{A}\bar{P}_m(A;\vec{b})\bar{F}^{(m)}(x),
\end{eqnarray}
where the passing-through term is omitted. Eq.(19) agrees with the result in [12]. Similarly, taking the $<\bar{M_L}|\ldots |\bar{N}>$ matrix element of Eq.(18), we get 
\begin{eqnarray*}
\rho_{M_L}^{\bar{N}A}(x)=\int d\vec{b}\sum_{l=0}^{A-1}\sum_{s=0}^{l}\sum_{t=0}^{A-l-1}\lambda_a(\vec{b})P_{s}^{L}(l;\vec{b})\bar{P}_{t}(A-l-1;\vec{b}) 
\end{eqnarray*}
\begin{eqnarray}
\ \ \ \ \times \int_{x}^{1}\frac{dy_1}{y_1}\int_{y_1}^{1}\frac{dy_2}{y_2}\int_{y_2}^{1}\frac{dy_3}{y_3} L^{(s)}(\frac{x}{y_1}) A(\frac{y_1}{y_2})\bar{F}^{(t)} (y_3),
\end{eqnarray}
where $L^{(s)}(x)$ and $\bar{F}^{(t)} (x)$  are defined in a similar form of Eq.(13). From Eq.(19), we get
\[\sigma_{non}^{\bar{N}A}=\int \frac{dx}{x}\rho_{\bar{N}}^{\bar{N}A}(x)=\int d\bar{b}[(1-\lambda_a(\vec{b}))^A-(1-\lambda_n(\vec{b})-\lambda_a(\vec{b}))^A].\]

(2) $\bar{N}+A \to N+X$ and $M_L+X$  for the target fragmentation regions($x<0$)

\vspace{0.3cm}
We apply the sum decomposition rule of Eq.(53) in Appendix to $\hat{\mbox{\boldmath $Q$}}_{\bar{N}A}(\vec{b},\alpha)$ 
>From Eq.(15), we have
\begin{eqnarray}
{\bf Spc} \hat{\mbox {\boldmath $Q$}}_{\bar{N}A}^{{\rm I}}(\vec{b},\alpha)=A(1-\lambda_a(\vec{b}))^{A-1}\hat{Q}_n(\vec{b},\alpha).
\end{eqnarray}
 From Eq.(16), we have
\begin{eqnarray*}
{\bf Spc} \hat{\mbox {\boldmath $Q$}}_{\bar{N}A}^{{\rm II}}(\vec{b},\alpha)=\frac{1}{\lambda_a(\vec{b})}[1-(1-\lambda_a(\vec{b}))^{A}-\lambda_a(\vec{b})A(1-\lambda_a(\vec{b}))^{A-1}]\hat{Q}_n(\vec{b},\alpha)
\end{eqnarray*}
\begin{eqnarray}
+[1-(1-\lambda_a(\vec{b}))^{A}]\hat{A}(\alpha)+\frac{1}{\lambda_a(\vec{b})}[(1-\lambda_a(\vec{b}))^{A-1}+A\lambda_a(\vec{b})-1]\hat{Q}_L(\vec{b},\alpha).
\end{eqnarray}

Taking the $<N|\ldots |N>$ matrix element of Eqs.(21) and (22) except for the passing-through terms, we have 
\begin{eqnarray*}
\rho_{N}^{\bar{N}A}(x)=\int d\vec{b}\{ \frac{\lambda_n(\vec{b})}{\lambda_a(\vec{b})} [1-(1-\lambda_a(\vec{b}))^A] \bar{F}(x)
\end{eqnarray*}
\begin{eqnarray}
+\frac{\lambda_L(\vec{b})}{\lambda_a(\vec{b})}[(1-\lambda_a(\vec{b}))^{A-1}+A\lambda_a(\vec{b})-1]N(x) \}.
\end{eqnarray}
While we take the $<M_L|\ldots |N>$ matrix element of Eq.(22), we get
\begin{eqnarray}
\rho_{M_L}^{\bar{N}A}(x)=\int d\vec{b}\ [1-(1-\lambda_a(\vec{b}))^{A}]A(x).
\end{eqnarray}
 From Eq.(24), we get 
\[\sigma_{an}^{\bar{N}A}=\int_0^1\frac{dx}{x}\rho_{M_L}^{\bar{N}A}(x)=\int \vec{b}[1-(1-\lambda_a)^A].\]
Thus, we obtain $\sigma_{inel}^{\bar{N}A}=\sigma_{non}^{\bar{N}A}+\sigma_{an}^{\bar{N}A}=\int d\vec{b}[1-(1-\lambda_n(\vec{b})-\lambda_a(\vec{b}))^A]$.

\vspace{0.5cm}

\small{\section{ Single-particle distributions in $\bar{A}$-$A$ collisions}}

\small{\subsection{Example($\bar{D}$-$D$ collisions)}} 

We consider $\bar{D}$-$D$ collisions with $\bar{A}=A=2$ in order to show the gross features of $\bar{A}$-$A$ collisions. All mechanisms in the operator set((a),(b),(c) and (d)) contribute to $\bar{A}$-$A$ collisions. In the following discussions, we omit  the arguments $\vec{b}$ and $\alpha$ for simplicity. We consider the projectile fragmentation regions of $\bar{D}$$(x>0)$. We obtain the seven mechanisms as follows;

(1) Non-annihilation interaction term
\begin{eqnarray}
 {\bf Spr}\left[
                            \matrix{
                            \hat{Q}_n & \hat{Q}_n \cr
                             \hat{Q}_n & \hat{Q}_n \cr
                             }
                             \right ] =2(1-\lambda_a)^2\hat{Q}_n^2.
\end{eqnarray}

(2) Annihilation terms of order $\lambda_a$
\begin{eqnarray}
{\bf Spr}\left[
                            \matrix{
                            \hat{A} & \hat{Q}_n \cr
                             \hat{Q}_n & \hat{Q}_n \cr
                             }
                             \right ] =\lambda_a[(1-\lambda_a)^2\hat{A}\hat{Q}_n+(1-\lambda_a)\hat{Q}_n^2],
\end{eqnarray}
\begin{eqnarray}
 {\bf Spr}\left[
                            \matrix{
                            \hat{Q}_L & \hat{A} \cr
                             \hat{Q}_n & \hat{Q}_n \cr
                             }
                             \right ] =\lambda_a[(1-\lambda_a)^2\hat{Q}_L\hat{A}+\hat{Q}_n^2],
\end{eqnarray}
\begin{eqnarray}
 {\bf Spr}\left[
                            \matrix{
                            \hat{Q}_L & \hat{Q}_n \cr
                             \hat{A} & \hat{Q}_n \cr
                             }
                             \right ] =\lambda_a(1-\lambda_a)[\hat{Q}_L\hat{Q}_n+\hat{A}\hat{Q}_n],
\end{eqnarray}
\begin{eqnarray}
 {\bf Spr}\left[
                            \matrix{
                            \hat{Q}_n & \hat{Q}_L \cr
                             \hat{Q}_L & \hat{A} \cr
                             }
                             \right ] =\lambda_a[\hat{Q}_n\hat{Q}_L+(1-\lambda_a)\hat{Q}_L\hat{A}].
\end{eqnarray}

(3) Annihilation terms of order $\lambda_a^2$
\begin{eqnarray}
 {\bf Spr}\left[
                            \matrix{
                            \hat{A}  & \hat{Q}_L \cr
                             \hat{Q}_L & \hat{A} \cr
                             }
                             \right ] =\lambda_a^2[\hat{A}\hat{Q}_L+\hat{Q}_L\hat{A}],
\end{eqnarray}
\begin{eqnarray}
 {\bf Spr}\left[
                            \matrix{
                            \hat{M}_0 & \hat{A} \cr
                             \hat{A} & \hat{Q}_n \cr
                             }
                             \right ] =\lambda_a^2[(1-\lambda_a)\hat{M}_0\hat{A}+\hat{A}\hat{Q}_n],
\end{eqnarray}
where the relations  $\eta_L+\lambda_L=1$ and $\eta_m+\lambda_m=1$ are used.
The diagram for Eq.(31) in $\bar{D}$-$D$ collision is shown in Fig.4.

\setlength{\unitlength}{1mm}
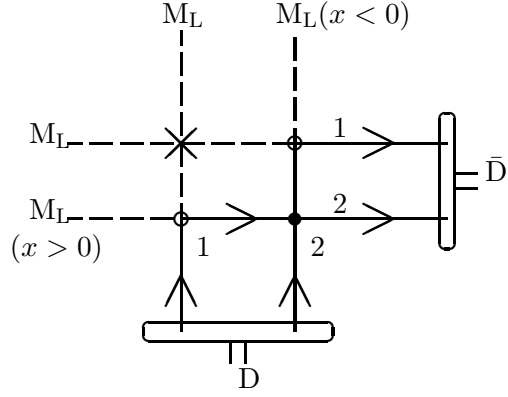
\begin{figure}
\begin{center}
\begin{picture}(120,40)
\thicklines

\put(60,35){\line(20,0){20}}
\put(45,25){\line(35,0){35}}

\put(30,35){\line(2,0){2}}
\put(33,35){\line(2,0){2}}
\put(34,35){\line(2,0){2}}
\put(37,35){\line(2,0){2}}
\put(38,35){\line(2,0){2}}
\put(41,35){\line(2,0){2}}
\put(42,35){\line(2,0){2}}
\put(45,35){\line(2,0){2}}
\put(46,35){\line(2,0){2}}
\put(49,35){\line(2,0){2}}
\put(50,35){\line(2,0){2}}
\put(53,35){\line(2,0){2}}
\put(54,35){\line(2,0){2}}
\put(57,35){\line(2,0){2}}
\put(58,35){\line(2,0){2}}
\put(60,35){\line(2,0){2}}

\put(30,25){\line(2,0){2}}
\put(33,25){\line(2,0){2}}
\put(34,25){\line(2,0){2}}
\put(37,25){\line(2,0){2}}
\put(38,25){\line(2,0){2}}
\put(41,25){\line(2,0){2}}
\put(42,25){\line(2,0){2}}
\put(45,25){\line(2,0){2}}

\put(45,10){\line(0,15){15}}
\put(60,10){\line(0,25){25}}

\put(45,25){\line(0,2){2}}
\put(45,28){\line(0,2){2}}
\put(45,29){\line(0,2){2}}
\put(45,32){\line(0,2){2}}
\put(45,33){\line(0,2){2}}
\put(45,36){\line(0,2){2}}
\put(45,37){\line(0,2){2}}
\put(45,40){\line(0,2){2}}
\put(45,41){\line(0,2){2}}
\put(45,44){\line(0,2){2}}
\put(45,45){\line(0,2){2}}
\put(45,48){\line(0,2){2}}

\put(60,35){\line(0,2){2}}
\put(60,38){\line(0,2){2}}
\put(60,39){\line(0,2){2}}
\put(60,42){\line(0,2){2}}
\put(60,43){\line(0,2){2}}
\put(60,46){\line(0,2){2}}
\put(60,47){\line(0,2){2}}

\put(73,35){\line(-2,1){4}}
\put(73,35){\line(-2,-1){4}}
\put(73,25){\line(-2,-1){4}}
\put(73,25){\line(-2,1){4}}

\put(45,17.5){\line(1,-2){2}}
\put(45,17.5){\line(-1,-2){2}}

\put(60,17.5){\line(1,-2){2}}
\put(60,17.5){\line(-1,-2){2}}

\put(55,25){\line(-2,-1){4}}
\put(55,25){\line(-2,1){4}}

\put(43,37){\line(1,-1){4}}
\put(43,33){\line(1,1){4}}
\put(60,35){\circle{1.7}}

\put(45,25){\circle{1.7}}
\put(60,25){\circle*{1.7}}

\put(65,35.8){1}
\put(65,25.8){2}

\put(47.0,20){1}
\put(62.0,20){2}

\put(25,35){${\rm M_L}$}
\put(25,25){${\rm M_L}$}

\put(42.5,51){${\rm M_L}$}
\put(57.5,51){${\rm M_L}$}

\put(80,30){\oval(2,18)}
\put(52.5,10){\oval(25,2.5)}

\put(81,31){\line(3,0){3}}
\put(81,29){\line(3,0){3}}
\put(51.5,8.75){\line(0,-3){3}}
\put(53.5,8.75){\line(0,-3){3}}

\put(63,51){$(x<0)$}
\put(22.5,20){$(x>0)$}

\put(85,30){$\bar{{\rm D}}$}
\put(52.5,2.5){D}

\end{picture}
\end{center}
\caption{Diagram for Eq.(31) in $\bar{D}$-$D$ collision. The intersection points represent $M_L$-$M_L$ collision($\times$), $\bar{N}$-$N$ collision($\bullet$) and $\bar{N}$-$N$ annihilation interaction($\circ$). The solid and dashed lines denote nucleon($N$) and leading meson($M_L$), respectively.}
\end{figure}

\vspace{0.5cm}
\small{\subsection{Generalization}} 

We investigate $\bar{A}$-$A$ collisions characterized by $ \hat{\mbox {\boldmath $Q$}}_{\bar{A}A}(\vec{b},\alpha)$ with A rows and A columns. 

($\alpha$) Non-annihilation term

The non-annihilation term in $\bar{A}$-$A$ collisions is characterized by $ \hat{\mbox {\boldmath $Q$}}_{\bar{A}A}^n(\vec{b},\alpha)$. Its elements $\hat{Q}^n_{ij}$ with $i=1,2,\ldots,A$  and $j=1,2,\ldots,A$ are given by
\begin{eqnarray}
\hat{Q}^n_{ij}=\hat{Q}_n,\ \ \ \ \ i=1,2,\ldots,A \ {\rm and}\  j=1,2,\ldots,A.
\end{eqnarray}
Then we get
\begin{eqnarray}
{\bf Spr}\hat{\mbox {\boldmath $Q$}}_{\bar{A}A}^n={\bf Spr}[\hat{Q}^n_{ij}]=A(1-\lambda_a)^{A^2-A}( \hat{Q} _n)^A.
\end{eqnarray}
Taking the $<\bar{N}|\ldots|\bar{N}>$ matrix element of Eq.(33) except for the passing-through term, we get the contribution of the non-annihilation term to the single-particle distribution of $\bar{A}+A \to \bar{N}+X(x>0)$ as follows;
\begin{eqnarray}
\rho^{\bar{A}A}_{\bar{N},n}(x)=\int d\vec{b}A\sum_{m=1}^{A^2}\bar{P}_{m}(A^2;\vec{b})\sum_{l=1}^{A}H(l;A^2,m,A)F^{(l)}(x).
\end{eqnarray}
This form is similar to Eq.(12) for $A+B \to N+X(x>0)$. From Eq.(34), the non-annihilation cross section is given by
\[\sigma_{non}^{\bar{A}A}=\frac{1}{A}\int d\vec{b}\sum_{}^{}\bar{P}_m(A^2;\vec{b})=\int d\vec{b}[(1-\lambda(\vec{b}))^{A^2}-(1-\lambda_n-\lambda_a(\vec{b}))^{A^2}].\]

($\beta$) Annihilation terms of order $\lambda_a$

If the annihilation operator $\hat{A}$ is at the $k$-th row and $l$-th column and its operator matrix is given by $\hat{\mbox {\boldmath $Q$}}_{\bar{A}A}^{a1}(k,l)$, the elements $\hat{Q}^{a1}_{ij}(k,l)$ with $i=1,2,\ldots,A$  and $j=1,2,\ldots,A$ of $\hat{\mbox {\boldmath $Q$}}_{\bar{A}A}^{a1}(k,l)$ are given by
\begin{eqnarray}
 \hat{Q}^{a1}_{ij}(k,l)=\left\{ 
                                \begin{array}{ll}
                                 \hat{A}, &   i=k\  {\rm and}\   j=l ,  \cr
                                 \hat{Q}_L, &   i=1,2,\ldots, k-1 \ {\rm with}\   j= l \ {\rm and}\   j=1,2,\ldots, l-1\cr
                                & {\rm  with}\  i= l, \cr
                                 \hat{Q}_n, &  {\rm otherwise .}  
                                 \end{array}
                                 \right . 
 \end{eqnarray}
 Then, from Eq.(52) in Appendix,  we get
\begin{eqnarray*}
{\bf Spr}\hat{\mbox {\boldmath $Q$}}_{\bar{A}A}^{a}=\lambda_a [ (k-1)(1-\lambda_a)^{\omega+2}(\hat{Q}_n)^{l-1}\hat{Q}_L(\hat{Q}_n)^{A-l}
\end{eqnarray*}
\begin{eqnarray}
+(1-\lambda_a)^{\omega+l+1}(\hat{Q}_L)^{l-1}\hat{A}(\hat{Q}_n)^{A-l}+(A-k)(1-\lambda_a)^{\omega+1}(\hat{Q}_n)^{A}],
\end{eqnarray}
where $\omega=A^2-A-k-l.$

If the annihilation term of order $\lambda_a$ is characterized by $\hat{\mbox {\boldmath $Q$}}_{\bar{A}A}^{a}$, we get  the relation $\hat{\mbox {\boldmath $Q$}}_{\bar{A}A}^{a1}=\sum_{k=1}^{A}\sum_{l=1}^{A}\hat{\mbox {\boldmath $Q$}}_{\bar{A}A}^{a1}(k,l)$.

>From the first term and third term in right-hand side of Eq.(36), the contribution of the annihilation term of order $\lambda_a$ to the single-particle distribution of $\bar{A}+A \to \bar{N}+X(x>0)$ is given by
\begin{eqnarray*}
\rho_{\bar{N},a1}^{\bar{A}A}(\vec{b},\alpha)=(1-\lambda_a)^{A^2-2A+2}[A+\frac{1}{\lambda_a}(1-\frac{1}{1-\lambda_a^A})]\sum_{l=1}^{A}\sum_{s=0}^{l-1}\sum_{t=0}^{A-l}\bar{P}_s(l-1;\vec{b})
\end{eqnarray*}
\begin{eqnarray*}
\times \bar{P}_t(A-l;\vec{b})\frac{1}{(1-\lambda_a)^l}(\bar{F}(\alpha))^s(\eta_L+\lambda_LN(\alpha))(\bar{F}(\alpha))^t+\frac{1}{\lambda_a^2}(1-\lambda_a)^{A^2-3A+2}
\end{eqnarray*}
\begin{eqnarray}
\times [1-(1-\lambda_a)^A][1+(A-1)(1-\lambda_a)^A-A(1-\lambda_a)^{A-1}]\sum_{m=1}^{A}\bar{P}_m(A;\vec{b})(\bar{F}(\alpha))^m,
\end{eqnarray}
where the passing-through term is omitted.

>From the second term in right-hand side of Eq.(36), the contribution of the annihilation term of order $\lambda_a$ to the single-particle distribution of $\bar{A}+A \to M_L+X(x>0)$ as follows;
\begin{eqnarray*}
\rho_{\bar{M_L},a1}^{\bar{A}A}( \vec{b},\alpha)=\sum_{k=1}^{A}\sum_{l=1}^{A}\lambda_a(1-\lambda_a)^{A^2-A-k+1}<M_L|(\hat{Q}_L)^{l-1}\hat{A}(\hat{Q}_{n})^{A-l}|\bar{N}>,
\end{eqnarray*}
\begin{eqnarray*}
=(1-\lambda_a)^{A^2-2A+1}[1-(1-\lambda_a)^A]\sum_{l=1}^{A}\sum_{s=0}^{l-1}\sum_{t=0}^{A-l}P^L_s(l-1;\vec{b})\bar{P}_t(A-l;\vec{b})
\end{eqnarray*}
\begin{eqnarray}
\times (L(\alpha))^sA(\alpha)(\bar{F}(\alpha))^t.
\end{eqnarray}

($\gamma$) Annihilation terms of order $\lambda_a^2$

In this case, the annihilation collision of order $\lambda_a^2$ is characterized by $ \hat{\mbox {\boldmath $Q$}}_{\bar{A}A}^{a2}(k,l;k^{\prime},l^{\prime})$ and $ \hat{\mbox {\boldmath $Q$}}_{\bar{A}A}^{a2}(k,l^{\prime};k^{\prime},l)$. The elements of $\hat{Q}^{a2}_{ij}(k,l;k^{\prime},l^{\prime})$ of $ \hat{\mbox {\boldmath $Q$}}_{\bar{A}A}^{a2}(k,l;k^{\prime},l^{\prime})$ are given by
\begin{eqnarray}
\hat{Q}^{a2}_{ij}(k,l;k^{\prime},l^{\prime})=\left\{ 
                                \begin{array}{ll}
                                 \hat{A}, & i=k, j=l\  {\rm and}\  i=k^{\prime}, j=l^{\prime},  \cr
                                 \hat{Q}_L, & i=1,2,\ldots,k-1\ {\rm with}\  j=l,\  j=1,2,\ldots,l-1\cr
                                       & \ {\rm with}\  i=k,i=1,2,\ldots,k^{\prime}-1\ {\rm with} \ j=l^{\prime} \cr
                                      & \ {\rm and}\  j=1,2,\ldots,l^{\prime}-1 {\rm with}\  i=k^{\prime}, \cr
                                 \hat{Q}_n, & {\rm otherwise, }
                                 \end{array}
                                 \right.
                                 \end{eqnarray}
where $k < k^{\prime}$ and $l < l^{\prime}$.
Then we get
\begin{eqnarray*}
{\bf Spr}\hat{\mbox {\boldmath $Q$}}_{\bar{A}A}^{a2}(k,l;k^{\prime},l^{\prime})=\lambda_a^2 [ (k-1)(1-\lambda_a)^{\Omega-l-l^{\prime}}(\hat{Q}_n)^{l-1}\hat{Q}_L(\hat{Q}_n)^{l^{\prime}-l-1}\hat{Q}_L(\hat{Q}_n)^{A-l^{\prime}}
\end{eqnarray*}
\begin{eqnarray*}
+(1-\lambda_a)^{\Omega-l^\prime+3}(\hat{Q}_L)^{l-1}\hat{A}(\hat{Q}_n)^{l^\prime-l-1}\hat{Q}_L(\hat{Q}_n)^{A-l^{\prime}}
\end{eqnarray*}
\begin{eqnarray*}
+ (k^\prime-k-1)(1-\lambda_a)^{\Omega+A-l-l^{\prime}+1}(\hat{Q}_n)^{l-1}\hat{Q}_L(\hat{Q}_n)^{l^{A-l^\prime}}
\end{eqnarray*}
\begin{eqnarray}
+(1-\lambda_a)^{\Omega-l+2}(\hat{Q}_L)^{l^\prime-1}\hat{A}(\hat{Q}_n)^{A-l^\prime}+(A-k^\prime)(1-\lambda_a)^{\Omega-l^\prime-l+2}(\hat{Q}_n)^{A}],
\end{eqnarray}
where $\Omega=A^2-A-k-k^{\prime}$.
While, the elements of $\hat{Q}^{a2}_{ij}(k,l^{\prime};k^{\prime},l)$ of $ \hat{\mbox {\boldmath $Q$}}_{\bar{A}A}^{a2}(k,l^{\prime};k^{\prime},l)$ are given by
\begin{eqnarray}
\hat{Q}^{a2}_{ij}(k,l^{\prime};k^{\prime},l)=\left\{ 
                                \begin{array}{ll}
                                 \hat{A}, & i=k, j=l^{\prime}\  {\rm and}\  i=k^{\prime}, j=l,  \cr
                                 \hat{M}_0, & i=k\  {\rm and}\  j=l,  \cr
                                 \hat{Q}_L, & i=1,2,\ldots,k^{\prime}-1(i \ne k)\ {\rm with}\  j=l , i=1,2,\ldots\cr
                                 & ,k-1 \ {\rm with}\ j=l^{\prime} ,\  j=1,2,\ldots,l^{\prime}(j \ne l) \cr
                                 & \ {\rm with}\  i=k , j=1,2,\ldots,l-1{\rm with}\  i=k^{\prime},\cr
                                 \hat{Q}_n, & {\rm otherwise, }
                                 \end{array}
                                 \right.
                                 \end{eqnarray}
where $k < k^{\prime}$ and $l < l^{\prime}$.
Then we get
\begin{eqnarray*}
{\bf Spr}\hat{\mbox {\boldmath $Q$}}_{\bar{A}A}^{a2}(k,l^{\prime};k^{\prime},l)=\lambda_a [ (k-1)(1-\lambda_a)^{\Omega-l-l^{\prime}+5}(\hat{Q}_n)^{l-1}\hat{Q}_L(\hat{Q}_n)^{l^{\prime}-l-1}\hat{Q}_L(\hat{Q}_n)^{A-l^{\prime}}
\end{eqnarray*}
\begin{eqnarray*}
+(1-\lambda_a)^{\Omega-l+3}(\hat{Q}_L)^{l-1}\hat{M}_0(\hat{Q}_L)^{l^{\prime}-l-1}\hat{A}(\hat{Q}_n)^{A-l^{\prime}}
\end{eqnarray*}
\begin{eqnarray*}
+(k^{\prime}-k-1)(1-\lambda_a)^{\Omega-l+4}(\hat{Q}_n)^{l-1}\hat{Q}_L(\hat{Q}_n)^{l^{A-l}}
\end{eqnarray*}
\begin{eqnarray}
+(1-\lambda_a)^{\Omega-l^{\prime}+3}(\hat{Q}_L)^{l-1}\hat{A}(\hat{Q}_n)^{A-l}+(A-k^{\prime})(1-\lambda_a)^{\Omega-l-l^{\prime}+3}(\hat{Q}_n)^{A}].
\end{eqnarray}
If the annihilation term of order $\lambda_a^2$ is characterized by $\hat{\mbox{\boldmath $Q$}}^{a2}_{\bar{A}A}$, we get 
\begin{eqnarray*}
\hat{\mbox{\boldmath $Q$}}^{a2}_{\bar{A}A}=\sum_{k=1}^{A-1}\sum_{l=1}^{A-1}\sum_{k^{\prime}=k+1}^{A}\sum_{l^{\prime}=l+1}^{A}[ \hat{\mbox {\boldmath $Q$}}_{\bar{A}A}^{a2}(k,l;k^{\prime},l^{\prime})+ \hat{\mbox {\boldmath $Q$}}_{\bar{A}A}^{a2}(k,l^{\prime};k^{\prime},l)].
\end{eqnarray*}

>From Eqs.(40) and (42), we may calculate the single-particle distributions of $\bar{N}$ and $M_L$ in the projectile fragmentation regions of $\bar{A}$ by means of the following relations: 
\begin{eqnarray*}
\rho_{\bar{N},a2}^{\bar{A}A}(\vec{b},\alpha)=<\bar{N}|{\bf Spr}\hat{\mbox{\boldmath $Q$}}^{a2}_{\bar{A}A}|\bar{N}>-<{\rm passing-through\  term}>,
\end{eqnarray*}
\begin{eqnarray*}
\rho_{M_L,a2}^{\bar{A}A}(\vec{b},\alpha)=<M_L|{\bf Spr}\hat{\mbox{\boldmath $Q$}}^{a2}_{\bar{A}A}|\bar{N}>.
\end{eqnarray*}

($\delta$) Annihilation terms of order $\lambda_a^A$

We consider the annihilation collision of order $\lambda_a^A$. We investigate the single-particle distribution of the leading meson $M_L.$ We estimate the following two cases:

(1) $ \hat{\mbox {\boldmath $Q$}}_{\bar{A}A}^{aA,{\rm I}}$ whose matrix elements $\hat{Q}^{aA,{\rm I}}_{ij}$ are defined by

\begin{eqnarray}
 \hat{Q}^{aA,{\rm I}}_{ij}=\left\{ 
                                \begin{array}{ll}
                                 \hat{A}, &   i= j , \cr
                                 \hat{Q}_L, &   i\ne j, \cr 
                                 \end{array}
                                 \right  .
 \end{eqnarray}
where $i=1,2,\ldots,A$ and $j=1,2,\ldots,A$.  
Then we get
\begin{eqnarray}
{\bf Spr}\hat{\mbox {\boldmath $Q$}}_{\bar{A}A}^{aA,{\rm I}}=\lambda_a^A\sum_{l=0}^{A-1} (\hat{Q}_L)^{l}\hat{A}(\hat{Q}_L)^{A-l-1}.
\end{eqnarray}
    
(2) $ \hat{\mbox {\boldmath $Q$}}_{\bar{A}A}^{aA,{\rm II}}$ whose matrix elements $\hat{Q}^{aA,{\rm II}}_{ij}$ are defined by
\begin{eqnarray}
 \hat{Q}^{aA,{\rm II}}_{ij}=\left\{ 
                                \begin{array}{ll}
                                 \hat{A}, &   i+j=A+1,  \cr
                                 \hat{M}_0, &   i=1,2,\ldots, A-1  \ {\rm and}\   j=1,2,\ldots, A-i, \cr
                                 \hat{Q}_n, &  i=2,3,\ldots,A \ {\rm and} \ j=A-i+2,\ldots,A. 
                                 \end{array}
                                 \right  . 
 \end{eqnarray}
                                 
Then we get
\begin{eqnarray}
{\bf Spr}\hat{\mbox {\boldmath $Q$}}_{\bar{A}A}^{aA,{\rm II}}=\lambda_a^A\sum_{l=0}^{A-1}(1-\lambda_a)^{\frac{1}{2}(A^2+A)-l-2}( \hat{M}_0)^{l}\hat{A}(\hat{Q}_n)^{A-l-1}.
\end{eqnarray}

Taking the $<M_L|\ldots |\bar{N}>$ matrix element of Eq.(44), we obtain
\begin{eqnarray*}
\rho_{M_L,aA,{\rm I}}^{\bar{A}A}(x)=\int d\vec{b}\sum_{l=0}^{A-1}\sum_{s=0}^{l}\sum_{t=0}^{A-l-1}\lambda_a(\vec{b})^A P_{s}^{L}(l;\vec{b})P^L_{t}(A-l-1;\vec{b})
\end{eqnarray*}
\begin{eqnarray}
\ \ \ \ \times \int_{x}^{1}\frac{dy_1}{y_1}\int_{y_1}^{1}\frac{dy_2}{y_2}\int_{y_2}^{1}\frac{dy_3}{y_3} L^{(s)}(\frac{x}{y_1}) A(\frac{y_1}{y_2})\bar{N}^{(t)} (y_3).
\end{eqnarray}
While we take the  $<M_L|\ldots |\bar{N}>$ matrix element of Eq.(46), we obtain
\begin{eqnarray*}
\rho_{M_L,aA,{\rm II}}^{\bar{A}A}(x)=\int d\vec{b}\sum_{l=0}^{A-1}\sum_{s=0}^{l}\sum_{t=0}^{A-l-1}\lambda_a(\vec{b})^A(1-\lambda_a)^{\frac{1}{2}(A^2+A)-l-2}P_{s}^{m}(l;\vec{b})
\end{eqnarray*}
\begin{eqnarray}
\ \ \ \ \times \bar{P}_{t}(A-l-1;\vec{b}) \int_{x}^{1}\frac{dy_1}{y_1}\int_{y_1}^{1}\frac{dy_2}{y_2}\int_{y_2}^{1}\frac{dy_3}{y_3} M_m^{(s)}(\frac{x}{y_1}) A(\frac{y_1}{y_2})\bar{F}^{(t)} (y_3),
\end{eqnarray}
where $P_s^m(l;\vec{b})=\left(\matrix{ l \cr
 s \cr}\right)\eta_m(\vec{b})^{l-s}\lambda_m({\vec{b})^s}$.

We obtain the similar results for the target fragmentation regions($x<0$).

\vspace{0.5cm}

\small{\section{Conclusion and Discussion }} 

We have investigated the single-particle distribution of $N$, $\bar{N}$ and $M_L$ in $A$-$A$ and $\bar{A}$-$A$ collisions on the basis of EMCM. By introduction of the operator matrix with the rule of the sum decomposition, a unified treatment of $A$-$A$ and $\bar{A}$-$A$ collisions at high energy is given. The analytic forms of the single-particle distribution of the inclusive process $A(\bar{A})+B \to C+X$ are derived. 

We have mainly discussed the peripheral reactions. The introduction of the non-peripheral reaction $M_L+N(\bar{N}) \to N(\bar{N})+M_L$+ hadrons gives the contributions of the higher order of $\lambda_a$ to the inclusive distributions of $\bar{A}$-$A$ collisions. Furthermore, the inclusive distributions of the secondary particles and the transverse momentum distribution are the unsolved problems. Also, the investigation of the relation to the Monte Carlo simulation models is one of the interesting problems in future.

\vspace{1cm}

{\bf Appendix }\ \ Definition and properties of operator matrix $\hat{\mbox {\boldmath $O$}}_{AB}(\alpha)$

\vspace{0.5cm}

The operator matrix $\hat{\mbox {\boldmath $O$}}_{AB}(\alpha)$ in the moment space is defined as the rectangular or square array of the operators $\hat{O}_{ij}(\alpha) $ with $A$ rows and $B$ columns 
\begin{equation}
\hat{\mbox {\boldmath $O$}}_{AB}(\alpha)=\left[
                                                \matrix{
                                                \hat{O}_{11}(\alpha) &  \hat{O}_{12}(\alpha) &  \cdots &  \hat{O}_{1B}(\alpha) \cr
                                                        \vdots & \vdots & \ddots & \vdots \cr
                                                 \hat{O}_{A1}(\alpha) &  \hat{O}_{A2}(\alpha) &  \cdots &  \hat{O}_{AB}(\alpha) \cr     
                                                 }
                                                 \right] .
\end{equation}

The operator matrix $\hat{\mbox {\boldmath $O$}}_{AB}(\alpha)$  has the following properties:

(i) The multiplication of $\hat{O}_{ij}(\alpha)$ by the scalar quantity(probability) $a_{ij}$ is defined by

\begin{eqnarray}
             \left[
                                                \matrix{
                                               a_{11} \hat{O}_{11}(\alpha)  &  \cdots & a_{1B} \hat{O}_{1B}(\alpha) \cr
                                                        \vdots & \ddots & \vdots \cr
                                                a_{A1} \hat{O}_{A1}(\alpha)  &  \cdots & a_{AB} \hat{O}_{AB}(\alpha) \cr     
                                                 }
                                                 \right]  =  \prod_{i=1}^{A}\prod_{j=1}^{B}a_{ij}
                                                 \left[
                                                  \matrix{
                                                \hat{O}_{11}(\alpha)  &  \cdots &  \hat{O}_{1B}(\alpha) \cr
                                                        \vdots & \ddots & \vdots \cr
                                                 \hat{O}_{A1}(\alpha)  &  \cdots &  \hat{O}_{AB}(\alpha) \cr     
                                                 }
                                                 \right] .
                                                   \end{eqnarray}

(ii) The rule of the operator addition is defined by

\begin{eqnarray*}
            \left[
                                                \matrix{
                                                \hat{O}_{11}(\alpha) + \hat{P}_{11}(\alpha) &  \hat{O}_{12}(\alpha) &  \cdots &  \hat{O}_{1B}(\alpha) \cr
                                                 \hat{O}_{21}(\alpha) &  \hat{O}_{22}(\alpha) &  \cdots &  \hat{O}_{2B}(\alpha) \cr 
                                                        \vdots & \vdots & \ddots & \vdots \cr
                                                 \hat{O}_{A1}(\alpha) &  \hat{O}_{A2}(\alpha) &  \cdots &  \hat{O}_{AB}(\alpha) \cr     
                                                 }
                                                 \right]   
                                                      \end{eqnarray*}
                                                 \begin{eqnarray}                                      
                                                =  \left[
                                                \matrix{
                                                \hat{O}_{11}(\alpha) &  \hat{O}_{12}(\alpha) &  \cdots &  \hat{O}_{1B}(\alpha) \cr
                                                 \hat{O}_{21}(\alpha) &  \hat{O}_{22}(\alpha) &  \cdots &  \hat{O}_{2B}(\alpha) \cr 
                                                        \vdots & \vdots & \ddots & \vdots \cr
                                                 \hat{O}_{A1}(\alpha) &  \hat{O}_{A2}(\alpha) &  \cdots &  \hat{O}_{AB}(\alpha) \cr     
                                                 }
                                                 \right]   
                                                 +
                                                \left[
                                                \matrix{
                                                \hat{P}_{11}(\alpha) &  \hat{O}_{12}(\alpha) &  \cdots &  \hat{O}_{1B}(\alpha) \cr
                                                 \hat{O}_{21}(\alpha) &  \hat{O}_{22}(\alpha) &  \cdots &  \hat{O}_{2B}(\alpha) \cr 
                                                        \vdots & \vdots & \ddots & \vdots \cr
                                                 \hat{O}_{A1}(\alpha) &  \hat{O}_{A2}(\alpha) &  \cdots &  \hat{O}_{AB}(\alpha) \cr     
                                                 }
                                                 \right] .   
\end{eqnarray}

(iii) The sum decomposition rule of $\hat{\mbox {\boldmath $O$}}_{AB}(\alpha)$

(a) The sum of the products of the row elements(${\bf Spr}  \hat{\mbox {\boldmath $O$}}_{AB}(\alpha)$)
\begin{eqnarray}
{\bf Spr} \hat{\mbox {\boldmath $O$}}_{AB}(\alpha) \equiv \sum_{i=1}^{A}\prod_{j=1}^{B}\hat{O}_{ij}(\alpha).
\end{eqnarray}

(b) The sum of the products of the column elements(${\bf Spc} \hat{\mbox {\boldmath $O$}}_{AB}(\alpha)$)
\begin{eqnarray}
{\bf Spc} \hat{\mbox {\boldmath $O$}}_{AB}(\alpha) \equiv \sum_{j=1}^{B}\prod_{i=1}^{A}\hat{O}_{ij}(\alpha).
\end{eqnarray}

The factorization rule of Eq.(50) is very important. From Eq.(50), we may calculate the event  probability in $A$-$B$ collisions correctly.

Notice that the sum decomposition rule must be used after the scalar factors $a_{ij}$ are taken outside. Namely, for Eq.(50), we have
\begin{eqnarray}
{\bf Spr}
             \left[
                                                \matrix{
                                               a_{11} \hat{O}_{11}(\alpha)  &  \cdots & a_{1B} \hat{O}_{1B}(\alpha) \cr
                                                        \vdots & \ddots & \vdots \cr
                                                a_{A1} \hat{O}_{A1}(\alpha)  &  \cdots & a_{AB} \hat{O}_{AB}(\alpha) \cr     
                                                 }
                                                 \right] \cr
                                                  =  \prod_{i=1}^{A}\prod_{j=1}^{B}a_{ij}{\bf Spr}
                                                 \left[
                                                  \matrix{
                                                \hat{O}_{11}(\alpha)  &  \cdots &  \hat{O}_{1B}(\alpha) \cr
                                                        \vdots & \ddots & \vdots \cr
                                                 \hat{O}_{A1}(\alpha)  &  \cdots &  \hat{O}_{AB}(\alpha) \cr     
                                                 }
                                                 \right] .
                                                   \end{eqnarray}
For Eq.(51), we have

\begin{eqnarray*}
{\bf Spr}
            \left[
                                                \matrix{
                                                \hat{O}_{11}(\alpha) + \hat{P}_{11}(\alpha) &  \hat{O}_{12}(\alpha) &  \cdots &  \hat{O}_{1B}(\alpha) \cr
                                                 \hat{O}_{21}(\alpha) &  \hat{O}_{22}(\alpha) &  \cdots &  \hat{O}_{2B}(\alpha) \cr 
                                                        \vdots & \vdots & \ddots & \vdots \cr
                                                 \hat{O}_{A1}(\alpha) &  \hat{O}_{A2}(\alpha) &  \cdots &  \hat{O}_{AB}(\alpha) \cr     
                                                 }
                                                 \right] 
                                                     \end{eqnarray*}
                                                 \begin{eqnarray}
                                                = {\bf Spr} \left[
                                                \matrix{
                                                \hat{O}_{11}(\alpha) &  \hat{O}_{12}(\alpha) &  \cdots &  \hat{O}_{1B}(\alpha) \cr
                                                 \hat{O}_{21}(\alpha) &  \hat{O}_{22}(\alpha) &  \cdots &  \hat{O}_{2B}(\alpha) \cr 
                                                        \vdots & \vdots & \ddots & \vdots \cr
                                                 \hat{O}_{A1}(\alpha) &  \hat{O}_{A2}(\alpha) &  \cdots &  \hat{O}_{AB}(\alpha) \cr     
                                                 }
                                                 \right]   \cr
                                                  +
                                                {\bf Spr}  \left[
                                                \matrix{
                                                \hat{P}_{11}(\alpha) &  \hat{O}_{12}(\alpha) &  \cdots &  \hat{O}_{1B}(\alpha) \cr
                                                 \hat{O}_{21}(\alpha) &  \hat{O}_{22}(\alpha) &  \cdots &  \hat{O}_{2B}(\alpha) \cr 
                                                        \vdots & \vdots & \ddots & \vdots \cr
                                                 \hat{O}_{A1}(\alpha) &  \hat{O}_{A2}(\alpha) &  \cdots &  \hat{O}_{AB}(\alpha) \cr     
                                                 }
                                                 \right] .   
\end{eqnarray}

Also, Eq.(53) satisfies the similar relations. According to these rules, it is easy to show the following relations for Eq.(49) with $\hat{O}_{ij}(\alpha)=\hat{Q}_0(\alpha)$ for $i=1,2,\ldots,A$ and $j=1,2,\ldots,B$ where $\hat{Q}_{0}(\alpha)=\eta\hat{G}(\alpha)+\lambda \hat{J}(\alpha)$:
\begin{eqnarray}
{\bf Spr} \left[
                                                \matrix{
                                                \hat{O}_{0}(\alpha) &  \hat{O}_{0}(\alpha) &  \cdots &  \hat{O}_{0}(\alpha) \cr
                                                        \vdots & \vdots & \ddots & \vdots \cr
                                                 \hat{O}_{0}(\alpha) &  \hat{O}_{0}(\alpha) &  \cdots &  \hat{O}_{0}(\alpha) \cr     
                                                 }
                                                 \right]  = A(\eta +\lambda) ^{AB-B} (\hat{Q}_0(\alpha))^B.
\end{eqnarray}

\end{document}